\documentclass[runningheads]{llncs}
\usepackage[T1]{fontenc}
\usepackage{graphics} 
\usepackage{epsfig} 
\usepackage{amssymb}  
\usepackage{amsmath} 
\usepackage{tabularx}
\usepackage{multirow}
\usepackage{booktabs}
\usepackage{multicol}
\usepackage{float}
\usepackage{svg}

\usepackage{array}

\begin{document}
\title{End-to-end autoencoding architecture for the simultaneous generation of medical images and corresponding segmentation masks}
%
%
\author{Aghiles Kebaili\inst{1, *} \and
Jérôme Lapuyade-Lahorgue\inst{1} \and
Pierre Vera\inst{2} \and 
Su Ruan\inst{1}}
\authorrunning{A. Kebaili et al.}
%
\institute{University of Rouen-Normandy, LITIS - QuantIF UR 4108, F-76000 Rouen, France \and
Department of Nuclear Medicine, Henri Becquerel Cancer Center, Rouen, France\\ \inst{*}\email{aghiles.kebaili@univ-rouen.fr}}
\maketitle              
\begin{abstract}
Despite the increasing use of deep learning in medical image segmentation, acquiring sufficient training data remains a challenge in the medical field. In response, data augmentation techniques have been proposed; however, the generation of diverse and realistic medical images and their corresponding masks remains a difficult task, especially when working with insufficient training sets. To address these limitations, we present an end-to-end architecture based on the Hamiltonian Variational Autoencoder (HVAE). This approach yields an improved posterior distribution approximation compared to traditional Variational Autoencoders (VAE), resulting in higher image generation quality. Our method outperforms generative adversarial architectures under data-scarce conditions, showcasing enhancements in image quality and precise tumor mask synthesis. We conduct experiments on two publicly available datasets, MICCAI's Brain Tumor Segmentation Challenge (BRATS), and Head and Neck Tumor Segmentation Challenge (HECKTOR), demonstrating the effectiveness of our method on different medical imaging modalities.

\keywords{Deep Learning  \and Data Augmentation \and Tumor Segmentation \and Generative Modeling \and Variational Autoencoder \and MRI \and PET.}
\end{abstract}
\section{Introduction}
Deep learning has made progress in medical imaging with successful outcomes in segmentation tasks \cite{isensee2021nnu,zhou2022missing} across various modalities, including Magnetic Resonance Imaging (MRI) \cite{lundervold2019overview} or Positron Emission Tomography (PET) \cite{islam2020gan}. However, the scarcity of medical imaging data poses a significant challenge, especially when physicians must manually delineate tumor masks, which can be tedious and time-consuming. Data augmentation is a popular technique in computer vision, which artificially generates new samples and can increase the size of the training set. Generative adversarial networks (GANs) \cite{goodfellow2014generative} have been widely used in medical imaging \cite{sandfort2019data} and recommended in multiple data augmentation literature reviews due to their ability to generate realistic images \cite{chen2022generative}. However, GANs have limitations such as learning instability, convergence difficulties, and mode collapse \cite{mescheder2018training} where the generator can only produce a limited range of possible samples. As an alternative, variational autoencoders (VAEs) \cite{kingma2013auto} have gained attention as a data augmentation approach that can outperform GANs in terms of output diversity and avoiding mode collapse. Recent research has proposed various methods to enhance VAE performance, given that VAEs tend to produce blurry and hazy output images due to the nature of the loss function \cite{kebaili2023deep}. Traditionally, conditional VAEs have been used for medical image augmentation \cite{zhuang2019fmri}, they allow for the control of the output samples by conditioning the generation process on additional informations. This approach enables the generation of synthetic samples that are representative of specific subgroups within the data and generating synthetic samples that are more targeted to specific tasks \cite{biffi2018learning}. Current studies \cite{gan2022esophageal,liang2021data} primarily address image generation or image transformation. In the context of increasing data for tumor segmentation, the generation of images alone is insufficient. It is crucial to not only generate images but also produce corresponding tumor masks, which serve as the ground truth for accurate segmentation. Few studies have explored the use of VAE-based architectures for the joint generation of both images and their corresponding tumor masks. Huo et al.'s work \cite{huo2022brain} is among the few that have explored this area. The method employed by the authors is centered around a progressive adversarial VAE architecture. Their approach, although innovative, is limited to generate only tumor regions with corresponding masks, which are then inserted directly into authentic images to produce synthetic samples. The masks in this context are given a priori and not generated during the process.

In this paper, we propose an approach for the simulataneos generation of medical images and corresponding tumor masks based on a Hamiltonian VAE (HVAE) \cite{caterini2018hamiltonian} architecture, which offers a better latent representation in comparison to the conventional VAE, as it incorporates Hamiltonian dynamics to encode complex data distributions. This method leads to a tighter bound on the log-likelihood, and can result in better realism of the generated images. Our proposed method is explicitly tailored for the task of tumor segmentation, as delineating tumors in medical images can be a tedious task. The simultaneous generation of images and their corresponding tumor masks can be an effective solution to increase the amount of data available for deep model training. To further elevate the efficacy of our architecture, we incorporate self-attention modules across deeper layers and the bottleneck, coupled with residual blocks for a better gradient flow.

\begin{itemize}
    \item Novel generative end-to-end architecture based on the HVAE for the simultaneous synthesis of medical images and corresponding segmentation masks.
    \item Robust model in the context of data-scarce scenarios.
    \item Demonstrated efficacy through a comprehensive segmentation evaluation on benchmark datasets: BRATS and HECKTOR. Our proposed approach showcases superior segmentation performance compared to contemporary GAN-based architecture.
\end{itemize}

\begin{figure*}[h]
\begin{center}
\includegraphics[width=11cm]{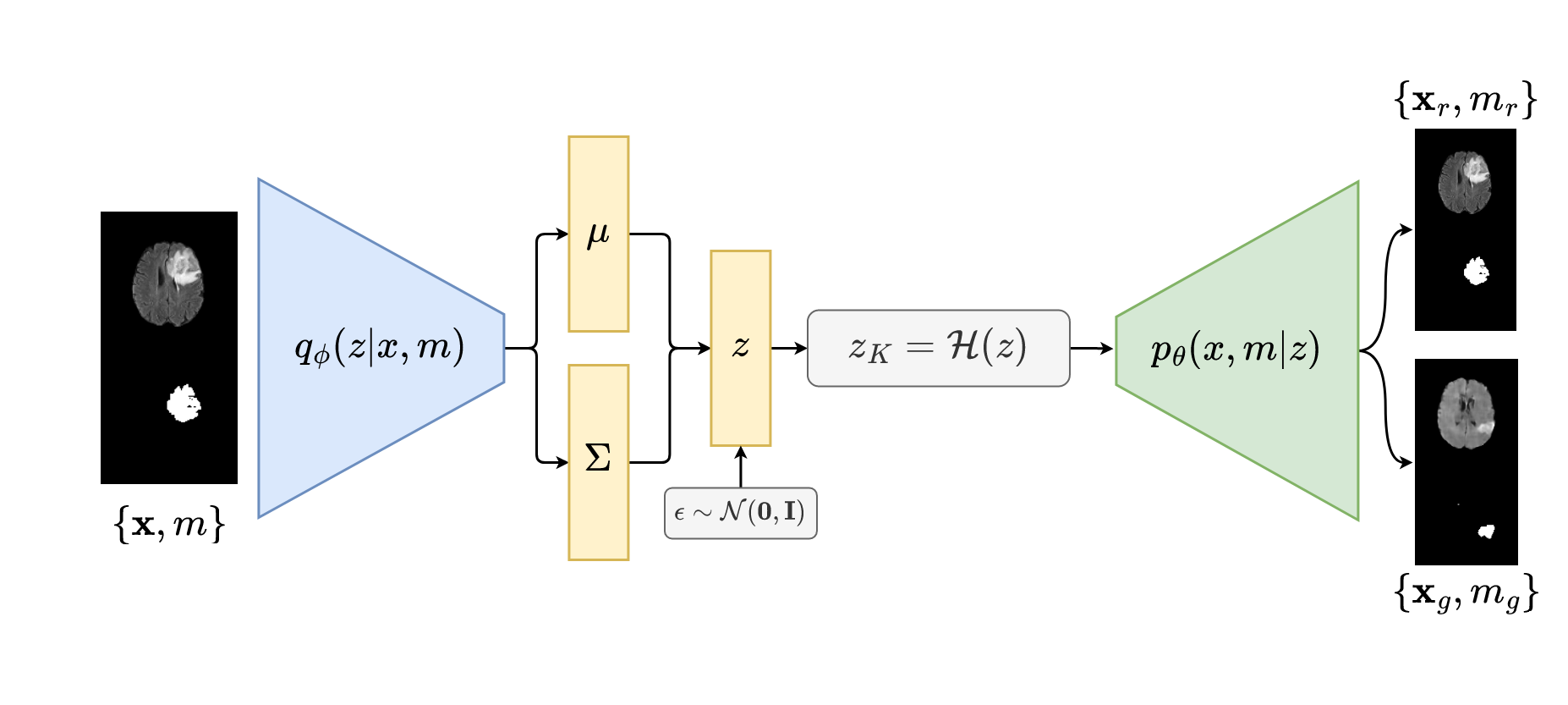}
\caption{Proposed architecture for medical image segmentation consists of an encoder $q_\phi$ and decoder $p_\theta$ network. It takes as concatenated multi-channel input the medical image and its corresponding tumor mask noted $\{\mathbf{x}, m\}$ and reconstructs it into $\{\mathbf{x}_r, m_r\}$ during training phase. Newly generated image pairs, denoted as ${\mathbf{x}_g, m_g}$, are produced during inference by feeding the decoder with a random Gaussian noise vector $z \sim \mathcal{N}(\mathbf{0}, \mathbf{I})$}.\label{fig1}
\end{center}
\end{figure*}

\section{Proposed method}
One advantage of the VAE approach is its ability to operate better with smaller datasets \cite{delgado2021deep} compared to GANs. This advantage can be attributed to the presence of an encoder in the VAE architecture, which extracts relevant features from input images for the generation of new ones. The encoder's ability to identify relevant features reduces significantly the search space required for generating new images through the process of reconstruction, unlike GANs, which have a wider search space and may encounter challenges in learning features effectively. In essence, VAEs for generation can be considered as a form of dimensionality reduction. The representation obtained by the encoder provides a better starting point for the decoder to accurately approximate the real data distribution. To this end, we propose an end-to-end HVAE-based architecture for generating medical images and their corresponding tumor masks. In the next section, we will provide a detailed explanation of our proposed architecture which is illustrated in Figure \ref{fig1}.

\subsection{The Hamiltonian VAE}
The HVAE is a modification of the standard VAE that allows for a better approximation of the posterior distribution $p_\theta(z|x)$ of the latent space $z$ given an input $x$, leading to better sampling quality and less blurry images. The HVAE is based on the Hamiltonian dynamics, which involves treating the latent variables as particles moving through a high-dimensional space according to the physical principles of energy conservation and momentum \cite{caterini2018hamiltonian}. More precisely, particles are generated according to the numerical approximation of the solution of Hamiltonian equations. Hamiltonian equations (\ref{eq2}) are two differential equations derived from the Hamiltonian defined by (\ref{eq1}). It was first introduced in physical mechanics for modeling the total energy in function of two variables: The position and the impulsion (motion quantity). The Hamiltonian equations describe how position and impulsion evolve in function of the time. The idea behind the HVAE is to introduce an auxiliary variable $\rho \sim \mathcal{N}(\mathbf{0}, \mathbf{M})$ called momentum, and use Hamiltonian dynamics for generation. The Hamiltonian in HVAE is defined as:

\begin{equation}
\mathcal{H}(z, \rho) = -\log p_\theta(z|x) + \frac{1}{2} \rho^T \mathbf{M}^{-1} \rho ,\label{eq1}
\end{equation}
where the first term is the potential energy and the second is the kinetic energy. The HVAE employs a Hybrid Monte Carlo sampling technique to leverage the Hamiltonian dynamics and explore the posterior distribution more effectively. The idea is to sample $(z, \rho)$ using the dynamic, this leads to the creation of an ergodic and time-reversible Markov chain for $z$ for which the stationary distribution matches the target distribution. $\rho$ and $z$ are updated using :

\begin{equation}
\frac{\partial \rho}{\partial t}(t) = -\nabla_z \mathcal{H}, \;\;\;\;\; \frac{\partial z}{\partial t}(t) = \nabla_\rho \mathcal{H}, \label{eq2}
\end{equation}
where $t$ is the step the of Markov chain. The details of the differentiable discretization of these two equations in (2) can be found in \cite{caterini2018hamiltonian}. A final proposal $z_K$, where $K$ is the total number of steps, is generated and accepted using the acceptance rejection algorithm with the Metropolis-Hastings ratio \cite{caterini2018hamiltonian}, which is used to balance between the exploration of the new state $z_K$ and the exploitation of the current state $z_{K-1}$.

Similarly to the vanilla VAEs, the HVAE is also based on a variational inference technique that estimates the posterior distribution $p_\theta(z|x)$ using a simpler distribution $q_\phi(z|x)$ through the maximization of the evidence lower bound term (ELBO). The posterior and variational distribution parameters are denoted by $\theta$ and $\phi$ respectively. In the case of the HVAE, the loss function is denoted as $\mathcal{L}_{\text{ELBO}}^{\text{H}}$. Building upon the Hamiltonian Monte Carlo (HMC) method, we introduce additional analytical evaluations for some terms within the expectation \cite{caterini2018hamiltonian}. Specifically, we compute the loss function as follows:

\begin{equation}
    \mathcal{L}_{\text{ELBO}}^{\text{H}} = \mathbf{E}_{z_0 \sim q_{\theta, \phi}} [\log p_\theta(x, z_K) - \frac{1}{2}\rho_K^T \rho_K - \log q_{\theta, \phi}(z_0)] ,\label{eq3}
\end{equation}
where $q_{\theta, \phi}(z_0) \sim \mathcal{N}(\mathbf{0}, \mathbf{I})$ is the log-likelihood of the initial latent vector $z_0$ over the Gaussian distribution.

\subsection{Modeling the joint distribution for the simultaneous medical images and masks generation}
Throughout the training process, our proposed generative models are conditioned on tumor masks by directly concatenating these masks with their corresponding medical images, creating a multi-channel input format. This enables us to effectively capture the joint distribution of images and masks, facilitating their concurrent generation during inference. By modeling the joint distribution, the generative model can simultaneously consider the dependencies between the images $x$ and the mask $m$, leading to a more precise and realistic image generation. While an alternative approach involves conditioning the model at the latent space level, a more conventional practice, we opted for the former method to avoid a two-stage procedure as proposed in \cite{huo2022brain,guibas2017synthetic}. The traditional two-stage process involves initially generating masks independently and subsequently utilizing them as conditions for the synthesis of medical images. By integrating the masks into our model, we are able to reformulate our loss function to model the joint distribution as follows :
\begin{equation}
\mathcal{L}_{\text{ELBO}}^{\text{H}} = \mathbb{E}_{z_0 \sim q_{\phi}} [\log p_\theta(x, z_K, m) - \frac{1}{2}\rho_K^T \mathbf{M}^{-1} \rho_K\\ - \log q_{\phi}(z_K|x, m)] \label{eq4}
\end{equation}
the term $\log p_\theta(x, z_K, m)$ can be further decomposed into two distinct components: $\log p_\theta(x, m | z_K)$ and $\log p_\theta(z_K)$. The first component, $\log p_\theta(x, m | z_K)$, denotes the reconstruction process for both the image and its associated mask, given on the latent vector $z_K$. To adress the mask reconstruction, we enhance the loss function by introducing an extra term that measures the cross-entropy between the reconstructed mask and the actual ground truth mask. Conversely, the second element, denoted as $\log p_\theta(z_K)$, defines the logarithmic probability of the vector $z_K$.

\section{EXPERIMENTS}
In this study, we conducted a comprehensive comparative analysis, evaluating our proposed architecture alongside the vanilla VAE architecture \cite{kingma2013auto} and the LSGAN \cite{mao2017least} as data augmentors for the tumor segmentation task. This research underscores the performance of our proposed generator in terms of image and mask diversity, underscoring its effectiveness in accommodating limited datasets. Evaluation was performed on established datasets: BRATS \cite{baid2021rsna} and HECKTOR \cite{andrearczyk2022overview}. Model performance was quantified using the Dice similarity coefficient (DSC) for segmentation. Additionally, image quality was assessed using Peak Signal-to-Noise Ratio (PSNR) and Structural Similarity (SSIM). The summarized outcomes are presented in Tables 1 and 2.

\subsection{Datasets}
We assess the effectiveness of our proposed approach through the utilization of two extensively employed publicly accessible datasets: the BRAin Tumor Segmentation (BRATS2021) dataset and the HEad and neCK tumOR segmentation (HECKTOR2022) dataset. These datasets encompass two diverse medical imaging modalities, specifically MRI and PET scans.
The BRATS dataset comprises 1258 subjects with MRI scans and segmentation mask including 3 types of tumor labels. The dimensions of MR images in the BRATS dataset are 240 $\times$ 240 \cite{baid2021rsna} and a voxel resolution of 1$\times$1$\times$1 $mm^3$. The HECKTOR dataset contains approximately 882 subjects including a PET and CT scans for each patient, and its corresponding segmentation mask. The PET scans have variable and lower dimensions, to ensure their consistency, we used tools from TorchIO to standardize and align the entire dataset to 256 $\times$ 256 with a voxel resolution of 1$\times$1$\times$1 $mm^3$ around the tumors. The tumor masks are binary masks \cite{andrearczyk2022overview}. In this study, we exclusively employ the FLAIR modality from the BRATS dataset and only PET modality from the HECKTOR dataset to test the efficacy of our proposed method.

\subsection{Training settings}
U-Net \cite{ronneberger2015u} is the most commonly used network for medical image segmentation. We thus choose it for demonstrate the performance of our method. We observed notable improvements in the segmentation task assessment using a U-Net architecture \cite{ronneberger2015u}, achieving satisfactory DSC scores of approximately 0.833$\pm$0.02 with around 300 images from the BRATS dataset and 0.741\%$\pm$0.03 with around 500 images from the HECKTOR dataset. To simulate real-world scenarios with limited data, we deliberately constrained the training dataset to merely 100 randomly selected samples from the entire patient cohort for both datasets. Evaluation was performed on a distinct test set of 100 samples. Two distinct learning phases were conducted. The first phase involved training the generative models with the authentic dataset, while the second phase encompassed training the segmentation models with the augmented datasets. The U-Net model was trained for a fixed duration of 80 epochs with a batch size of 32 in each experimental setup. The mean DSC and its corresponding standard deviation were computed across 10 independent runs, enhancing the validation of results, especially given the limited dataset scenario where result variability could be high.

\newcolumntype{P}[1]{>{\centering\arraybackslash}p{#1}}
\begin{table}[t]
\caption{Quantitative performance of the generative models is evaluated in terms of DSC(\%↑) on BRATS and HECKTOR.}\label{tab1}
\begin{center}
\begin{tabular}{P{3cm}P{3cm}P{3cm}P{3cm}}
\toprule
\multicolumn{2}{c}{Methods} & BRATS                       & HECKTOR                     \\ \cmidrule(r){1-2} \cmidrule(lr){3-3} \cmidrule(l){4-4}
\multicolumn{2}{c}{Reference (real image)}                                                                       & 0.633$\pm$0.14 & 0.442$\pm$0.09 \\
 \cmidrule(r){1-2} \cmidrule(lr){3-3} \cmidrule(l){4-4}
\multicolumn{2}{c}{Augmented ($\times$2)}                                                                              & 0.740$\pm$0.05 & 0.584$\pm$0.06 \\
\multicolumn{2}{c}{Augmented ($\times$3)}                                                                              & 0.772$\pm$0.05 & 0.639$\pm$0.03 \\
\multicolumn{2}{c}{Augmented ($\times$5)}                                                                              & 0.800$\pm$0.18 & 0.668$\pm$0.01 \\
 \cmidrule(r){1-2} \cmidrule(lr){3-3} \cmidrule(l){4-4}
\multicolumn{1}{c}{\multirow{4}{*}{LSGAN}}                                                    & + 100 synthetic & 0.680$\pm$0.03 & 0.576$\pm$0.11 \\
\multicolumn{1}{c}{}                                                                          & + 200 synthetic & 0.703$\pm$0.04 & 0.689$\pm$0.03 \\
\multicolumn{1}{c}{}                                                                          & + 300 synthetic & 0.685$\pm$0.02 & 0.680$\pm$0.02 \\
\multicolumn{1}{c}{}                                                                          & + 500 synthetic & 0.667$\pm$0.08 & 0.678$\pm$0.03 \\
 \cmidrule(r){1-2} \cmidrule(lr){3-3} \cmidrule(l){4-4}
\multicolumn{1}{c}{\multirow{4}{*}{VAE}} & + 100 synthetic & 0.751$\pm$0.01 & 0.570$\pm$0.05 \\
\multicolumn{1}{c}{} & + 200 synthetic & 0.760$\pm$0.01 & 0.662$\pm$0.05 \\
\multicolumn{1}{c}{} & + 300 synthetic & 0.752$\pm$0.04 & 0.702$\pm$0.02 \\
\multicolumn{1}{c}{} & + 500 synthetic & 0.726$\pm$0.06 & 0.712$\pm$0.01 \\
 \cmidrule(r){1-2} \cmidrule(lr){3-3} \cmidrule(l){4-4}
\multicolumn{1}{c}{\multirow{4}{*}{\begin{tabular}[c]{@{}c@{}}HVAE\\(ours)\end{tabular}}}                                                     & + 100 synthetic & 0.770$\pm$0.03 & 0.586$\pm$0.07 \\
\multicolumn{1}{c}{}                                                                          & + 200 synthetic & 0.781$\pm$0.02 & 0.659$\pm$0.04 \\
\multicolumn{1}{c}{}                                                                          & + 300 synthetic & \textbf{0.803$\pm$0.03} & 0.705$\pm$0.01 \\
\multicolumn{1}{c}{}                                                                          & + 500 synthetic & 0.760$\pm$0.09 & \textbf{0.713$\pm$0.01} \\ \bottomrule
\end{tabular}
\end{center}
\end{table}

\subsection{Evaluation results}
We first present the reference DSC obtained by training a U-Net on the baseline training set to demonstrate the difficulty to train the model with limited amount of data. Alternatively, traditional data augmentation techniques such as random rotation, flipping, and croppings were applied to the training set with augmentation factors of 2, 3, and 5, noted as \textit{Standard Aug.} ($\times$2), ($\times$3) and ($\times$5), respectively. Deep generative models including the two proposed architectures were then employed to generate synthetic images with corresponding tumor masks. Specifically, 100, 200, 300 and 500 synthetic images were generated, mixed with the 100 images from the real training set. The proposed data augmentation method was evaluated on the BRATS and HECKTOR datasets to demonstrate the model's ability to adapt to different imaging modalities. Results are depicted in Table \ref{tab1}.

Our proposed architecture improves the DSC on both the BRATS and HECKTOR datasets, demonstrating notable improvements of 26.8\% and 61.1\% respectively compared to the reference. Furthermore, the proposed method exhibits a low standard deviation (approximately 2\% DSC), affirming its robustness against data variability. Additionally, in terms of PSNR and SSIM, as depicted in table \ref{tab2}, our architecture outperforms state-of-the-art methods. On the BRATS dataset, it achieves a PSNR of 13.610 dB and an SSIM of 84.5\%, while on the HECKTOR dataset, it achieves a PSNR of 18.628 dB and an SSIM of 83.9\%. These evaluations were conducted using 100 synthetic and real images across 10 independent runs. It is also worth noting that our proposed architecture is the only method to surpass traditional augmentation techniques by a factor of 5, underscoring its superiority in enhancing dataset size and diversity.

\begin{figure*}[h]
\begin{center}
\includegraphics[width=12.5cm]{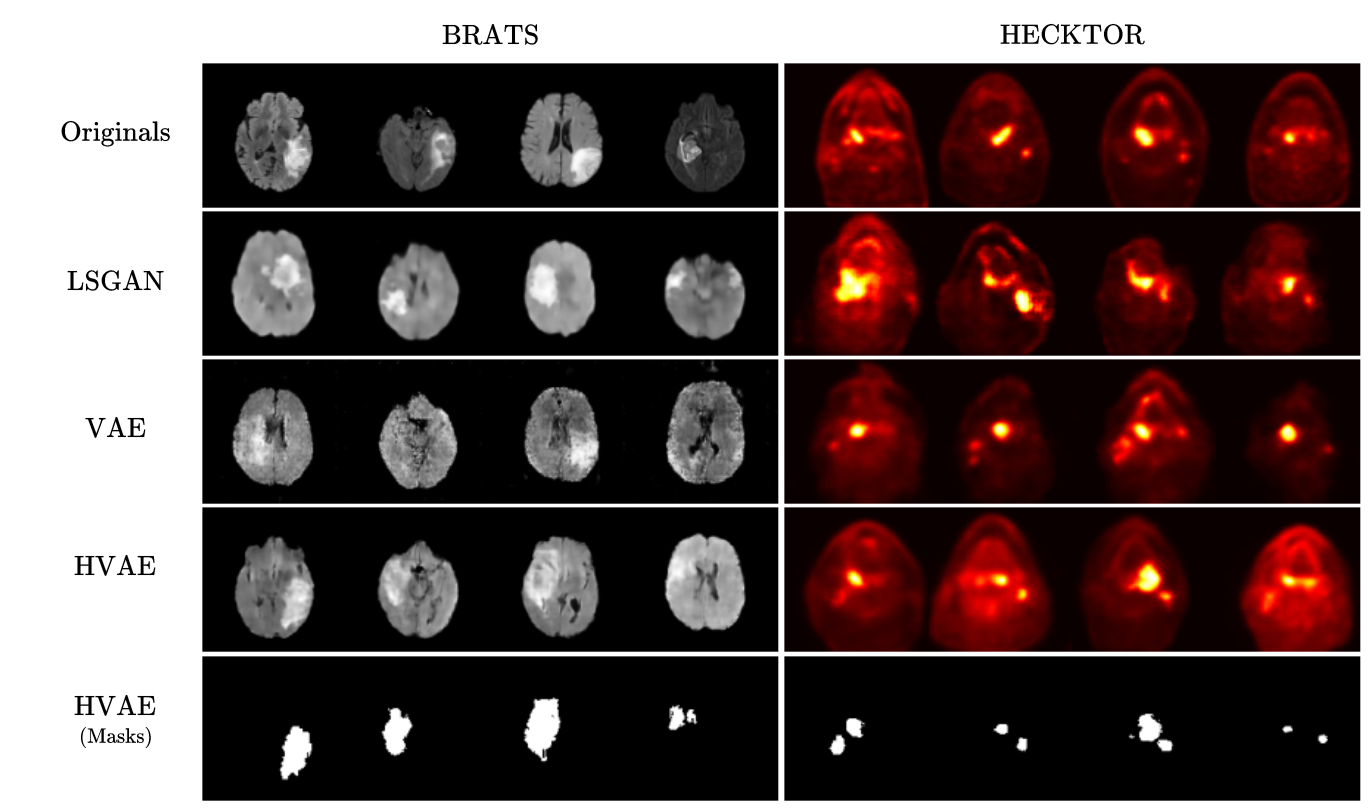}
\caption{Comparison of generated images of two state-of-the-art methods with our proposed ones through 4 examples represented in columns (HVAE with the generated tumor masks in the last row).} \label{fig2}
\end{center}
\end{figure*}

\begin{table}[h!]
\caption{{\small Quantitative performance of the generative models is evaluated in terms of PSNR (\textnormal{dB}↑) and SSIM (\%↑) between test set and generated set.}} \label{tab2}
\begin{center}
\begin{tabular}{P{2cm}P{2cm}P{2cm}P{2cm}P{2cm}}
\toprule
\multirow{2}{*}{Method} & \multicolumn{2}{c}{BRATS} & \multicolumn{2}{c}{HECKTOR} \\ \cmidrule(l){2-5} 
                        & PSNR (\textnormal{dB}↑) & SSIM (\%↑) & PSNR (\textnormal{dB}↑) & SSIM (\%↑) \\ \cmidrule(r){1-1} \cmidrule(lr){2-3} \cmidrule(l){4-5}
LSGAN                   & 12.956$\pm$0.04 & 0.612$\pm$0.02 & 16.130$\pm$0.02 & 0.726$\pm$0.03 \\
VAE                   & 13.272$\pm$0.07 & 0.837$\pm$0.02 & 17.966$\pm$0.03 & 0.826$\pm$0.04 \\
HVAE                   &  \textbf{13.610$\pm$0.03} & \textbf{0.845$\pm$0.01} & \textbf{18.628$\pm$0.05} & \textbf{0.839$\pm$0.05} \\ \bottomrule
\end{tabular}%
\end{center}
\end{table}

As depicted in Figure \ref{fig2}, we observe an enhancement in quality when transitioning from the VAE to the HVAE, resulting in a more realistic representation of tumor regions in MRI scans. Furthermore, the GAN architecture is particularly affected due to its high data requirements. These observations validate the hypothesis regarding the potential of VAEs and autoencoding architectures in dealing with insufficient datasets. Finally, we can note a decline in the DSC score beyond the inclusion of 500 synthetic images for the BRATS dataset, as indicated in Table \ref{tab1}. This decrease can be attributed to the imbalance between the number of real and synthetic images, introducing a bias that causes the model to primarily optimize the loss through synthetic images alone, negatively impacting the DSC score on the test set. However, it is worth mentioning that the DSC score starts decreasing beyond 200 synthetic MRIs for the LSGAN and vanilla VAE models, while this decrease occurs at more than 300 synthetic images for our archticture. This further demonstrates that the images generated by the HVAE are more realistic, and the U-Net architecture requires additional effort to perceive the perceptual difference between synthetic and real images.

\section{Conclusion}
In this study, we introduced a novel architecture based on the HVAE model for data augmentation in medical image segmentation tasks. The experimental results showed that our proposed method outperformed traditional data augmentation techniques and traditional generative adversarial models when trained under 3D data-scarce regimes . Our future directions include further improving our architecture into a hybrid model that combines strenghts of variational inference and adversarial learning, we will also potentually explore in mode details the modeling of the latent space as a Riemannian manifold.

%
%
%
%

\end{document}